\newcommand{\diracslash}[1]{#1\llap{/\kern2pt}}

\def\bearr{\begin{eqnarray}}

\def\eearr{\end{eqnarray}}

\newcommand{\be}{\begin{equation}}

\newcommand{\ee}{\end{equation}}

\newcommand{\bea}{\begin{eqnarray}}

\newcommand{\eea}{\end{eqnarray}}

\newcommand{\ba}[1]{\begin{array}{#1}}

\newcommand{\ea}{\end{array}}

\newcommand{\eqrf}[1]{Eq.\ (\ref{#1})}

\newcommand{\eqrftw}[2]{Eqs.\ (\ref{#1}) and (\ref{#2})}

\documentclass[preprint,secnumarabic,floats,amssymb,bibnotes,footinbib,tightenlines]{revtex4}

\usepackage{graphicx}
\usepackage{fixltx2e}
\usepackage{amsmath}
\usepackage{amssymb}
\usepackage{tabularx}

\usepackage{latexsym}

\usepackage{epsfig}

\usepackage{subfigure}
\usepackage{hyperref}
\usepackage[outdir=./]{epstopdf}

\begin{document}
\title{Observational constraints on $\alpha$-Starobinsky inflation}
\author{Saisandri Saini and Akhilesh Nautiyal}
\affiliation{Department of Physics, Malaviya National Insitute of Technology Jaipur,
JLN  Marg, Jaipur-302017, India}

\begin{abstract}
In this work we revisit  $\alpha$-Starobinsky inflation, also know as $E$-model,
in the light of current CMB and LSS observations. The inflaton potential in the 
Einstein frame for this model contains a parameter $\alpha$ in the exponential, which 
alters the predictions for the scalar and tensor power spectra of Starobinsky inflation. We obtain these power spectra 
  numerically without using slow-roll approximation and perform MCMC analysis  to 
put constraints on parameters $M$ and $\alpha$ from Planck-2018, BICEP/Keck (BK18) 
and other LSS observations. We consider general reheating scenario by varying the 
number of e-foldings during inflation, $N_{pivot}$,  along with the other 
parameters. We find $\log_{10}\alpha = 0.0^{+1.6}_{-5.6}$, 
$\log_{10}M= -4.91^{+0.69}_{-2.7}$
and $N_{pivot} = 53.2^{+3.9}_{-5}$ with $95\%$ C. L.. This implies that the 
present CMB and LSS observations are insufficient to constrain the 
parameter $\alpha$. We also find that there is no correlation between $N_{pivot}$ 
 and $\alpha$. 

\end{abstract}

\maketitle

\section{Introduction} Inflation \cite{Guth:1980zm} is a theoretical framework in 
cosmology that proposes a brief and extremely rapid  expansion of the 
early universe before big bang nucleosynthesis. It provides 
 solution to some key puzzles of the big bang model such as the horizon problem, 
which is related to the uniformity of the cosmic microwave background radiation, 
and the flatness problem, which concerns the geometry of the universe. 
The driving force behind  inflation is a hypothetical scalar field 
named as  inflaton, whose  potential energy dominates the energy density of the 
universe causing quasi-exponential expansion of the universe for a very short period 
of time. During this period the quantum fluctuations in the inflaton, which 
are coupled to the metric perturbations,  give rise to the primordial 
density perturbations. The quantum fluctuations in the spacetime geometry during
inflation are responsible for the primordial gravitational waves 
(tensor perturbations). These primordial perturbations generated during inflation
provide seeds for cosmic microwave background anisotropy and  structures in the 
universe. Observations of CMB anisotropy and polarization by 
COBE \cite{Smoot:1992td}, WMAP \cite{Komatsu:2010fb}, 
Planck \cite{Ade:2015lrj, Planck:2018jri}, and BICEP/Keck array offer robust 
experimental support for the predictions of inflation, i.e. adiabatic, nearly 
scale invariant and Gaussian perturbations. Since its inception, several models of 
inflation based on particle physics and string theory  have been explored 
\cite{Martin:2013tda,Martin:2024qnn}.  The most popular models of inflation, where inflaton has  quadratic or quartic potentials, have been ruled out by Planck 
observations \cite{Planck:2018jri}.

One of the well suited model of inflation from recent Planck and BICEP/Keck array
observations is Starobinsky inflation \cite{Starobinsky:1980te,Starobinsky:1983zz}
where inflation is achieved by $\frac{1}{M^2}R^2$ interaction, R being the Ricci 
scalar , in the Einstein Hilbert action without additional scalar field. 
In the Einstein frame the $R^2$ term gives rise to plateau potential of the 
inflaton, named as scalaron in this case. Starobinsky inflation is of great 
importance in the light of recent observations as it predicts a scalar tilt value of 
\cite{Mukhanov:1981xt} $n_s \simeq 0.965$ and a smaller value of 
tensor-to-scalar ratio $r\simeq 0.003$. Besides this, it also incorporates a
graceful exit to the radiation dominated epoch via reheating 
\cite{Vilenkin:1985md,Mijic:1986iv, Ford:1986sy}, where the standard 
model particles were produced by the oscillatory decay of scalaron. 

Another interesting feature of Starobinsky inflation
is that the scalaron potential in the Einstein frame can 
be easily realized in the framework of no-scale 
supergravity \cite{Ellis:2013xoa} with noncompact $SU(2,1)/SU(2)\times U(1)$ symmetry. In this case we have a modulus field
that can be fixed by the other dynamics \cite{Ellis:2013nxa}, and the inflaton field  is a part of chiral superfield
with a minimal Wess-Zumino superpotential. No-scale supergravity
\cite{Cremmer:1983bf,Ellis:1983sf,Lahanas:1986uc}, where the supersymmetry breaking scale is undetermined in a first 
approximation and the energy scale of the effective potential can be much smaller than the Planck scale, 
provides a framework to connect inflation to a viable quantum theory of gravity at high scales and the standard model of 
particle physics at lower scale. Attempts have been made to incorporate standard model
of particle physics in no-scale supergravity models of 
inflation \cite{Ellis:2013nka,Ellis:2015kqa,Ellis:2016ipm,Ellis:2017jcp}. 
In \cite{Ellis:2013nxa,Ellis:2018zya} various possible 
examples in no-scale supergravity framework, which 
can reproduce the effective potential of the Starobinsky
model and other related models, have been explored.

In this work we consider a variant of Starobinsky model known as $\alpha-$Starobinsky model 
of inflation \cite{Ellis:2013nxa}, where the Einstein
frame potential of the scalaron is modified by a parameter  $\alpha$ in the exponential.  
This potential is  obtained by generalizing the coefficient of the logarithm of the volume modulus field in the 
K\"ahler potenial in $SU(2,1)/SU(2)\times U(1)$ no-scale supergravity framework \cite{Ellis:1983ei,Ellis:1984bm} with 
a suitable choice of the superpotential having both the volume modulus field and the chiral superfield; the inflaton
being the scalar part of the chiral superfield. Similar potentials can also be obtained in supergravity models where the 
inflaton is a part of the vector multiplet rather than the chiral multiplet \cite{Ferrara:2013rsa,Kallosh:2013yoa}. 
In this case the $\alpha-$Starobinsky model belongs to a class of superconformal inflationary models known 
as $\alpha-$attractors. The parameter $\alpha$ is crucial as it is connected to various physical aspects, such as the geometry of
 the k$\ddot{a}$hler manifold, hyperbolic geometry \cite{Carrasco:2015rva,Kallosh:2015zsa,Carrasco:2015uma}, the behavior of 
the boundary of moduli space \cite{Kallosh:2014rga}, modified gravity \cite{Odintsov:2016vzz}, maximal supersymmetry 
\cite{Kallosh:2017ced}, and string theory \cite{Scalisi:2018eaz,Kallosh:2017wku}. 
In \cite{Alho:2017opd} $\alpha$-attractor models have been studied in the dynamical system framework
to find inflationary attractor solutions in the light of observational viability of these models. 
A two-field inflation model where both the fields have $\alpha$-attractor potentials is studied in \cite{Rodrigues:2020fle}.
The $E$-model potential in brane inflation along with its various observational aspects has been studied 
in \cite{Sabir:2019wel}. The primordial black hole formation in $\alpha$-attractor inflation models has been studied is  
\cite{Dalianis:2018frf},\cite{Mahbub:2019uhl} . 
  
The predictions for tensor-to-scalar ratio $r$ are modified in the $\alpha$-Starobinsky model by a factor 
of $\alpha$ \cite{Ellis:2013nxa,Kallosh:2013yoa}.  To obtain constraints on $\alpha$, this model was analyzed with Planck-2015 
observations in \cite{Planck:2015sxf}, where the power spectra was calculated using  public code ASPIC \cite{Martin:2013tda} based on Houble flow functions $\epsilon_i$. By choosing a flat prior over [0,4] for $\log_{10}(\alpha^2)$ it was found that 
$\log_{10}(\alpha^2)<1.7(2.0)$ for E model and $\log_{10}(\alpha^2)<2.3(2.5)$ at $95\%$ CL for T model 
of $\alpha$ attractors. Further, the Planck 2018 results have also placed an upper 
limit on the parameter 
$\alpha$ \cite{Planck:2018jri}, $\log_{10}\alpha<1.3$ $(\alpha<19)$ for the E-model and $\log_{10}\alpha<1.0$ $(\alpha<10)$ for the T-model 
at $95\%$ CL by choosing  the parameter range $-2 < \log_{10}\alpha < 4$. The constraints on $\alpha$ from Planck-2018 
observations for few choices of e-folding along with various phenomenological aspects of  this model were studied 
in \cite{Ellis:2020xmk}, and it was found that the data was insufficient to constrain $\alpha$. 
However, an upper limit on $\alpha$ ($\alpha\le 46 (88)$ for e-folds $N =50 (60)$) 
was obtained from the Planck upper 
limit on $r$ in \cite{Ueno:2016dim}. In \cite{Ellis:2021kad} the joint constraints 
on $n_s-r$ from Planck-2018 and 
BICEP/Keck observations were used to put constraints on the parameter $\alpha$ and the bounds on reheating temperature were 
used to put constraints on the e-folds $N$. Both these analysis were based on slow-roll approximation. 
Reheating constraints on $\alpha$-attractor models were also studied in  \cite{Ueno:2016dim} and it was found that 
 the parameter $\alpha$ is roughly constrained as $\alpha \ge 0.01$ along a broad resonance preheating 
scenario. By assuming a two-phase reheating with a preheating for specific duration, constraint on $\alpha$
has been obtained in \cite{ElBourakadi:2022lqf} and it is shown that small values of $\alpha$ give good results on $r$.
The constraint on parameter $\alpha$ has also been obtained in \cite{Sarkar:2021ird} by solving the cosmological
perturbation equations in $k$-space. Observational constraints on other variants of Statobinsky inflation has
been studied in \cite{SantosdaCosta:2020dyl,Saini:2023uqc} and it is shown that the current data allows small deviation from the Starobinsky potential in the Einstein frame.

Here, we analyze $\alpha$-Starobinsky inflation model in the light of 
Planck-2018, BICEP/Keck (BK18) and BAO observations to obtain constraints on 
parameter $\alpha$. To obtain the power spectra of primordial perturbations we use ModeChord \cite{2021ascl}, 
an updated version of ModeCode \cite{Mortonson:2010er}, which
solves the background and perturbation equations for inflaton numerically without the usual slow-roll approximation. 
ModeCode is coupled with CAMB \cite{Lewis:1999bs}, which computes the angular power spectra for CMB anisotropy and polarization.
The theoretical angular power spectra obtained with CAMB are used to compute constraints on inflationary parameters  
from CMB observations with the help of CosmoMC \cite{Lewis:2002ah}.  CosmoMC 
performs the Markov Chain Monte Carlo (MCMC) analysis, which is a statistical technique used to sample probability 
distributions of model parameters. By using ModeCode one can directly constrain the parameters of inflaton potential and 
$N_{pivot}$ (the number of e-foldings between horizon exit of the CMB pivot mode and the end of inflation) from CMB observations. 
To find the best-fit parameters of the $\alpha$-Starobinsky model, we vary $log_{10}\alpha$ between $-8.0$ to $4.0$ along with 
$M$ and $N_{pivot}$. 

The format of this paper is as follows: A brief discussion of $\alpha$ - Starobinsky model is given in Sec.\ref{alphastaromodel}. In section \ref{basicequation}, we 
obtain the background equations for Hubble parameter and inflaton in terms of e-foldings $N$. We also
find  perturbation equations for Mukhanov-Sasaki variable and tensor mode in terms of $N$. These equations are
 used in ModeCode. 
The details of MCMC analysis performed using CosmoMC along with observational constraints are described in 
section \ref{observconstra}.  In Section \ref{conclusions}, we provide our conclusions and a summary of our findings. 

\section{$\alpha$-Starobinsky model(E-model)} \label{alphastaromodel}
The action for the Starobinsky inflation \cite{Starobinsky:1980te,Starobinsky:1983zz}
 contains  $~R^2$ term due to quantum effects, and is given as

\begin{equation}
S_{J} = \frac{-M_{Pl}^2}{2}\int \sqrt{-g} (R + \mu R^2) d^4 x. \label{SJ} 
\end{equation}
Here $\mu = \frac{1}{6M^2}$,  $M$ is a parameter having a mass dimension of one
and $M << M_{Pl}$. $M_{Pl}$ is the reduced Planck mass ( $M_{Pl}^2 = (8\pi G)^{-1}$) 
and $g$ is the determinant of the metric $g_{\mu \nu}$. After this point, we will 
work in units where $M_{Pl}=1$.
After Weyl transformation $\tilde{g}_{\mu \nu}=(1+2\mu \phi)g_{\mu \nu}(x)$ 
and using the field redefinition $\chi \equiv \sqrt{\frac{3}{2}} \ln (1+\frac{\phi}{3M^2})$, the action \eqrf{SJ} gets transformed to an Einstein Hilbert form:
\begin{equation}
S_E = \frac{1}{2}\int d^4x\sqrt{-\tilde{g}}\left(-\tilde{R}+\partial_{\mu}{\chi}\partial^{\mu}{\chi}+\frac{3}{2}M^2(1-e^{-\sqrt\frac{2}{3}\chi})^2\right).\label{SE}
\end{equation}
It is evident from the above equation  that the inflaton potential  for Starobinsky inflation is  given by:
\begin{equation}
V(\chi) = \frac{3}{4}M^2(1-e^{-\sqrt\frac{2}{3}\chi})^2. \label{pot}
\end{equation}
This model's predictions for $n_s$ and r can be expressed as a function of the number of e-foldings $N_e$. In the limit of large $N_e$, one discovers
\begin{align}
n_s = 1-\frac{2}{N_e},
r = \frac{12}{N_e^2}.
\end{align}
It is shown in \cite{Ellis:2013xoa} that in the framework of   
no-scale supergravity with noncompact $SU(2,1)/SU(2)\times U(1)$ symmetry, one can obtain the inflaton potential (\ref{pot}),  
which can be generalized with the introduction of a new parameter $\alpha$ \cite{Ellis:2013nxa}.  This generalization is based
on the K\"ahler potential \cite{Ellis:1983ei,Ellis:1984bm} 	
\be
K=-3\alpha\ln\left(T+T^\dagger-\frac{\phi\phi^\dagger}{2}\right). \label{kahlerpot}
\ee 
Here the field $T$ corresponds to a generic compactification volume modulus and $\phi$ is a part of generic chiral mater field.
The parameter $\alpha$ is known as K\"ahler curvature parameter as it 
is inversely related to the curvature of the K\"ahler manifold ($R=\frac{2}{3\alpha}$). In \cite{Ellis:2013xoa} the  Wess-Zumino superpotential 
$W = M\left(\frac{\phi^2}{2} - \frac{\phi^3}{3\sqrt{3}} \right)$ was used to obtain the Starobinsky potential 
(\ref{pot}) in Einstein frame, along with the K\"ahler potential (\ref{kahlerpot}) for $\alpha=1$ yields
\be
V(\phi)= 3 M^2 \left(\frac{\phi}{\phi+\sqrt{3}}\right)^2, \label{kwpot}
\ee
which with field redefinition $\phi=\sqrt{3}\tanh\left(\frac{\chi}{\sqrt{6}}\right)$ gives the Starobinsky potential (\ref{pot}).

 For  $\alpha \ne 1$ the field redefinition can be generalized as \cite{Ellis:2019bmm} 
\be
\phi=\sqrt{3}\tanh\left(\frac{\chi}{\sqrt{6 \alpha}}\right).  
\ee
This, on substitution  in (\ref{kwpot}), yields the potential for $\alpha$-Starobinsky inflation as 
\begin{equation}
V(\chi) = \frac{3}{4}M^2(1-e^{-\sqrt\frac{2}{3\alpha}\chi})^2. \label{alphapot}
\end{equation}
For any value of $\alpha$ the potential (\ref{alphapot}) can also be generated by considering the 
superpotential \cite{Ellis:2019bmm} 
\be
W = \sqrt{\alpha}f(\phi)\left(2T-\frac{\phi^2}{3}\right)^{\frac{3}{2}(\alpha-\sqrt{\alpha})}. \label{superpotalpha}
\ee
To determine the function $f(\phi)$, $\phi$ and $T$ are considered to be real along with $\langle T \rangle = \frac{1}{2}$;
and the potential obtained from the superpotential (\ref{superpotalpha}) is equated to (\ref{kwpot}). The superpotential 
(\ref{superpotalpha}) can be successfully combined with the dark energy and supersymmetry breaking \cite{Ellis:2019bmm}.

\begin{figure}[h]
\begin{center}
\includegraphics[width=12cm, height = 7cm]{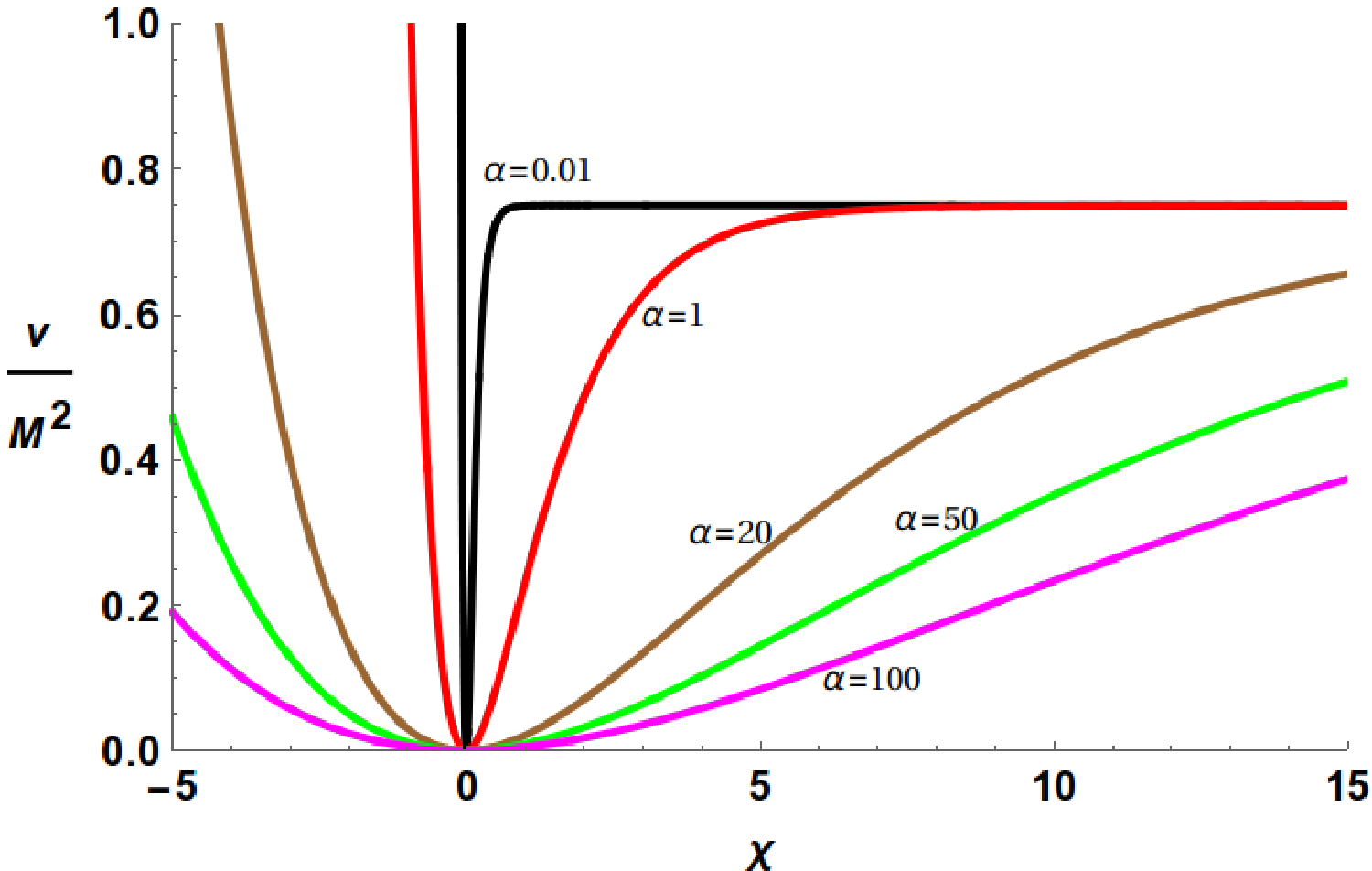}
\caption{The variation of potential (\ref{alphapot}) with $\alpha=0.01, 1, 20, 50, 100$. The red line corresponds to the original Starobinsky inflationary potential with $\alpha = 1$. The values of potential and field being in $M_{Pl} = 1$ units.}
\label{fig:plot}
\end{center}
\end{figure}

The potential (\ref{alphapot}) can also be derived from supergravity models where the inflaton belongs to the vector
superfield \cite{Ferrara:2013rsa,Kallosh:2013yoa}, and it belongs to a class of attractor models of inflation known as $\alpha$ - 
attractors. This potential, for $\alpha \le {\cal O}(1)$ and large $N$, leads to a general prediction for inflationary 
observables \cite{Ellis:2013nxa,Kallosh:2013yoa},
\begin{align}
n_s = 1-\frac{2}{N_e},
r = \frac{12\alpha}{N_e^2}.
\end{align}
For $\alpha = 1$, the potential (\ref{alphapot})  corresponds to the Starobinsky-Whitt potential (\ref{pot}).
 On the other hand, for $\sqrt{\frac{2}{3\alpha}}<<1,$ this potential is quadratic.
\be
V(\chi) = \frac{3}{4}M^2(1-e^{-\sqrt\frac{2}{3\alpha}\chi})^2 = \frac{m^2}{2}\phi^2,
\ee
where $m^2 = \frac{M^2}{2}$.
The  $\alpha$ - Starobinsky potential (\ref{alphapot}) for different value of $\alpha$ is plotted in
Fig.~\ref{fig:plot}. It is evident from the Fig. that  smaller the values of  $\alpha$, narrower the potential minima. 
The potential has a wall for tiny negative $\chi$ and has an inflationary plateau at large positive $\chi$. 
It can be seen from the figure that increasing the value of curvature parameter $\alpha$ stretches the Starobinsky potential 
horizontally, which reduces the flatness of the plateau at any fixed value of $\chi$. 
 
\section{Inflationary dynamics and power spectra} \label{basicequation}
 As mentioned earlier, we use ModeCode to obtain the power spectra of scalar and tensor perturbations generated during inflation.
ModeCode solves both the background and perturbation equations in terms of e-folds ($N$) as an independent variable 
numerically without slow-roll approximation. In this section we obtain the necessary equations describing the background 
dynamics during inflation and the perturbation equations in terms of e-folds. 
The evolution of the Hubble parameter during inflation is governed by the  Friedmann equations:  
\bea
H^2 &=& \frac{1}{3}\left[\frac{1}{2}\dot{\chi^2} + V(\chi)\right]. \label{H2}\\
\dot H &=& -\frac{1}{2 }\dot{\chi^2},  \label{hprime}
\eea
where $V(\chi)$ is the potential (\ref{alphapot}) for the inflaton $\chi$ in the Einstein frame. The dynamics of the scalar 
field  $\chi$  is governed by its equation of motion, which is the Klein-Gordon equation in spacetime: 
\be
\ddot{\chi} + 3H\dot{\chi} + \frac{dV(\chi)}{d\chi} = 0. \label{evo}
\ee
Here, the overdot denotes derivative with respect to cosmic time. 
As we use the number of e-folding,  $N = \ln a $ as an
independent variable to solve the equations numerically, we can rewrite the Friedmann equations (\ref{H2}) and (\ref{hprime}) 
and the  equation of motion of inflaton (\ref{evo})  in terms of $N$ as
\bea
H^2 &=& \frac{\frac{1}{3}V(\chi)}{1-\frac{1}{6}\chi^{\prime 2}},\label{H2n}\\
H^\prime &=& -\frac{1}{2}H\chi^{\prime 2},\label{Hprimen}
\eea
and 
\be
\chi^{\prime \prime}+\left(\frac{H^\prime}{H}+3\right)\chi^\prime+\frac{1}{H^2}\frac{dV(\chi)}{d\chi} = 0,\label{evon}
\ee
where prime denotes the differentiation with respect to $N$. The numerical solution of these equations 
provides various background quantities that are  used in perturbations equations.
 We describe the scalar perturbation generated during inflation  by the gauge invariant Mukhanov-Sasaki variable 
$u_k$ \cite{Mukhanov:1988jd,Sasaki:1986hm}, which is related to the curvature perturbations $\mathcal{R}$ as 
$u_k = -z \cal{R}$, where $z=\frac{1}{H}\frac{d\chi}{d\tau}$ and $\tau$ denotes the conformal time. 
The quantity $z$ can be determined from background equations \eqrftw{Hprimen}{evon}.
The Fourier modes 
\be
u_k(\tau) = \int d^3xe^{-ik.x}u(\tau,x), 
\ee
satisfies the Mukhanov-Sasaki equation
\be
\frac{d^2 u_k}{d\tau^2} + \left(k^2 - \frac{1}{z}\frac{d^2z}{d\tau^2}\right)u_k = 0,\label{uk}
\ee
The primordial power spectrum of scalar perturbations is given in terms of the two point correlation function of 
comoving curvature  and its relation with Mukhanov-Sasaki variable $u_k$  is given by \eqrf{Prs}and \eqrf{Ps} respectively:
\be
\mathcal{P_\mathcal{R}} = \frac{k^3}{2\pi^2}\langle \mathcal{R}_{k}\mathcal{R}_{k^\prime}^*\rangle\delta^3(k-k^\prime), 
\label{Prs}
\ee
which can be expressed in terms of Mukhanov-Sasaki variable $u_k$ as
\be
\mathcal{P_\mathcal{R}}(k) = \frac{k^3}{2\pi^2}\bigg|\frac{u_k}{z}\bigg|^2. \label{Ps}
\ee

Similarly, for tensor perturbations, the mode equation and the primordial tensor power spectrum is given by following equations:
\be
\frac{ d^2 v_k}{d\tau^2} + \left(k^2 - \frac{1}{a}\frac{d^2 a}{d\tau^2}\right)v_k = 0,\label{vk}
\ee
\be
\mathcal{P}_{t}(k) = \frac{4k^3}{\pi^2}\bigg|\frac{v_k}{a}\bigg|^2.\label{Pt}
\ee
To obtain the scalar and tensor power spectra, the mode equations (\ref{uk}) and (\ref{vk}) are solved numerically along with background equations (\ref{Hprimen}) and (\ref{evon}).
For this we can rewrite these equations in terms of  e-foldings $N=\ln a$ as

\bea
&u_k^{\prime \prime} + \left(\frac{H^\prime}{H}+1\right)u_k^{\prime}+\Biggl\{\frac{k^2}{a^2H^2}-\left[2-4\frac{H^\prime}{H}\frac{\chi^{\prime \prime}}{\chi^\prime}-2\left(\frac{H^\prime}{H}\right)^2 -5\frac{H^\prime}{H}-\frac{1}{H^2}\frac{d^2V}{d\chi^2}\right] \Biggr\}u_k = 0,\label{ukn}&\\
&v_k^{\prime \prime} + \left(\frac{H^\prime}{H}+1\right)v_k^{\prime} + \left[\frac{k^2}{a^2H^2} -\left(\frac{H^\prime}{H}+2\right) \right]v_k = 0.&\label{vkn}
\eea
In this approach the scalar spectral index $n_s$ and the tensor spectral index $n_t$ are not the parameter of the 
power spectra, and hence they are determined from the power spectra obtained numerically using their definitions 
\cite{Bassett:2005xm}:
\bea
n_s &=& 1 + \frac{d \ln\mathcal{P_\mathcal{R}}}{d \ln k},\label{ns}\\
n_t &=& \frac{d \ln\mathcal{P}_{t}}{d \ln k},\label{nt}
\eea
Similarly, the tensor-to-scalar ratio $r$,  defined by \cite{Bassett:2005xm}
\be
r = \frac{\mathcal{P}_{t}}{\mathcal{P_\mathcal{R}}},\label{r}
\ee
is also a derived parameter and is obtained from the numerical solution of the power spectra. Both $n_s$ and $r$ are determined
at the pivot scale $k=0.05$Mpc\textsuperscript{-1}.
By obtaining the power spectra numerically  the parameters of the inflaton potential (\ref{alphapot}), $M$ and $\alpha$ can be 
directly constrained from the CMB and LSS observations.

\section{OBSERVATIONAL CONSTRAINTS} \label{observconstra}
The background and perturbations equations, described in the previous section, are solved by ModeCode \cite{Mortonson:2010er}
numerically without using slow roll approximation. Thus, slow-roll violating effects, which can be significant for confronting 
models with precision data, are automatically captured. 
We modify ModeChord, un updated version of ModeCode,  to compute the scalar and tensor power spectra 
for $\alpha$-Starobinsky inflation in the Einstein frame. Inflationary models are described in ModeCode using an array of 
parameters for the potential. We incorporate the parameters $M$ and $\alpha$ of the potential (\ref{alphapot}) in this array. 
Without depending on the slow roll approximation, we consider the general reheating scenario, where the 
parameter $N_{pivot}$  representing the number of e-foldings 
from the end of inflation to the time when length scales corresponding to  the Fourier mode $k_{pivot}$ leave the 
horizon during inflation, is also varied along with other potential parameters.  The numerically computed primordial 
power spectra with ModeChord can be used in CAMB \cite{Lewis:1999bs}, which solves Boltzmann equation and computes the 
two-point correlation function  for CMB temperature anisotropy and polarization for a given set of cosmological parameters. 
CosmoMC uses these theoretical angular power spectra to put constraints on the parameters of the inflaton potential, 
$N_{pivot}$, and the other parameters of the $\lambda$CDM model from various CMB and large-scale structure observations. 
We use Planck-2018, BICEP (BK18) \cite{BICEP:2021xfz}, BAO and Pantheon data to 
determine the constraints on the parameters $M$
 and $\alpha$ of inflaton potential (\ref{alphapot}) and $N_{pivot}$ . We have used flat priors for these parameters, 
which are  given in Table \ref{Tab:prior}. To cover a wide range, we sample $M$ and $\alpha$ on a logarithmic scale. 
We also vary the parameters of the $\Lambda$CDM model with their priors provided in \cite{Planck:2018vyg}. 
To ensure MCMC convergence, we analyze four chains using the Gelman and Rubin $R-1$ statistics, which evaluates 
the ``variance of the mean" against the ``mean of the chain variance." 
 
\begin{table}[htbp]
\centering
\small
\begin{tabular}{|c|c|}
\hline
\textbf{Parameter} & \textbf{Prior range} \\
\hline
$\mathbf{N_{\text{pivot}}}$ & [20, 90] \\
\hline
$\log_{10} M$ & [$-10.0$, $-1.0$] \\
\hline
$\log_{10} \alpha$ & [$-8.0$, $4.0$] \\
\hline
\end{tabular}
\caption{Priors on $N_{\text{pivot}}$ and model parameters.}
\label{Tab:prior}
\end{table}

Table \ref{Tab:constraint} shows the  constraints obtained from MCMC analysis for parameters of potential (\ref{alphapot}), 
the e-foldings $N_{pivot}$ and the derived parameters, $r$ and $n_s$.  

\begin{table}[htbp]
\centering
\small  
\setlength{\tabcolsep}{8pt}  
\renewcommand{\arraystretch}{1.3}  
\begin{tabular}{|c|c|c|c|}
\hline
\textbf{Parameter} & \textbf{68\% limits} & \textbf{95\% limits} & \textbf{99\% limits} \\
\hline
{\boldmath$N_{\text{pivot}}$} & $53.2^{+2.9}_{-1.5}$ & $53.2^{+3.9}_{-5.0}$ & $53.2^{+4.3}_{-7.2}$ \\
\hline
{\boldmath$\log_{10} M$} & $-4.91^{+0.54}_{-0.043}$ & $-4.91^{+0.69}_{-2.7}$ & $-4.91^{+0.72}_{-3.3}$ \\
\hline
{\boldmath$\log_{10} \alpha$} & $0.022^{+1.2}_{+0.023}$ & $0.0^{+1.6}_{-5.6}$ & $0.0^{+1.4}_{-7.3}$ \\
\hline
{\boldmath$n_s$} & $0.9644^{+0.0027}_{-0.0017}$ & $0.9644^{+0.0038}_{-0.0044}$ & $0.9644^{+0.0045}_{-0.0064}$ \\
\hline
{\boldmath$r$} & $0.0121^{+0.0073}_{-0.012}$ & $0.012^{+0.020}_{-0.016}$ & $0.012^{+0.028}_{-0.019}$ \\
\hline
\end{tabular}
\caption{Constraints on parameters of the potential, $r$, and $n_s$ using Planck 2018, BICEP/Keck (BK18), and BAO observations.}
\label{Tab:constraint}
\end{table}

 It can be seen from the Table~\ref{Tab:constraint} that the mean value of $\alpha$ along with $95\%\, C.L.$ is
\be
\log_{10} \alpha = 0.0^{+1.6}_{-5.6}. \label{alphavalue}
\ee
As the  limits  are larger than the mean value,  $\alpha=1$ is consistent with the 
Planck-2018 and BICEP/Keck (BK18) observations. So the present data is not 
sufficient to constrain the value of 
$\alpha$; a similar result was obtained in \cite{Ellis:2020xmk}. 
The $95\%$ upper limit obtained on $\alpha$ in our analysis, i.e. $\log_{10}\alpha\le 1.6$ is slightly larger than that obtained by Planck-2018 \cite{Planck:2018jri}. 
This difference arises as we have varied $M$ and $N_{pivot}$ along with $\alpha$.
However, the upper limit on $\alpha$ obtained in our analysis, i.e, 
$\alpha\le 39.81$ with $95\%$ C.L., is smaller than the one obtained 
in \cite{Ellis:2021kad}, and the lower limit, $\alpha \ge 2.51\time 10^{-6}$ with
$95\%$ C.L., is also smaller than the lower limit obtained in \cite{Ueno:2016dim}
using a broad resonance preheating.
The number of e-folding obtained for $\alpha$-Starobinsky model with $95\%\, C. L.$  is 
\be
N_{pivot}=53.2^{+3.9}_{-5.0},
\ee
which is sufficient to solve the horizon problem.

\begin{figure}[h!]
\begin{center}
\subfigure[]{
 \includegraphics[width=8cm, height = 5cm]{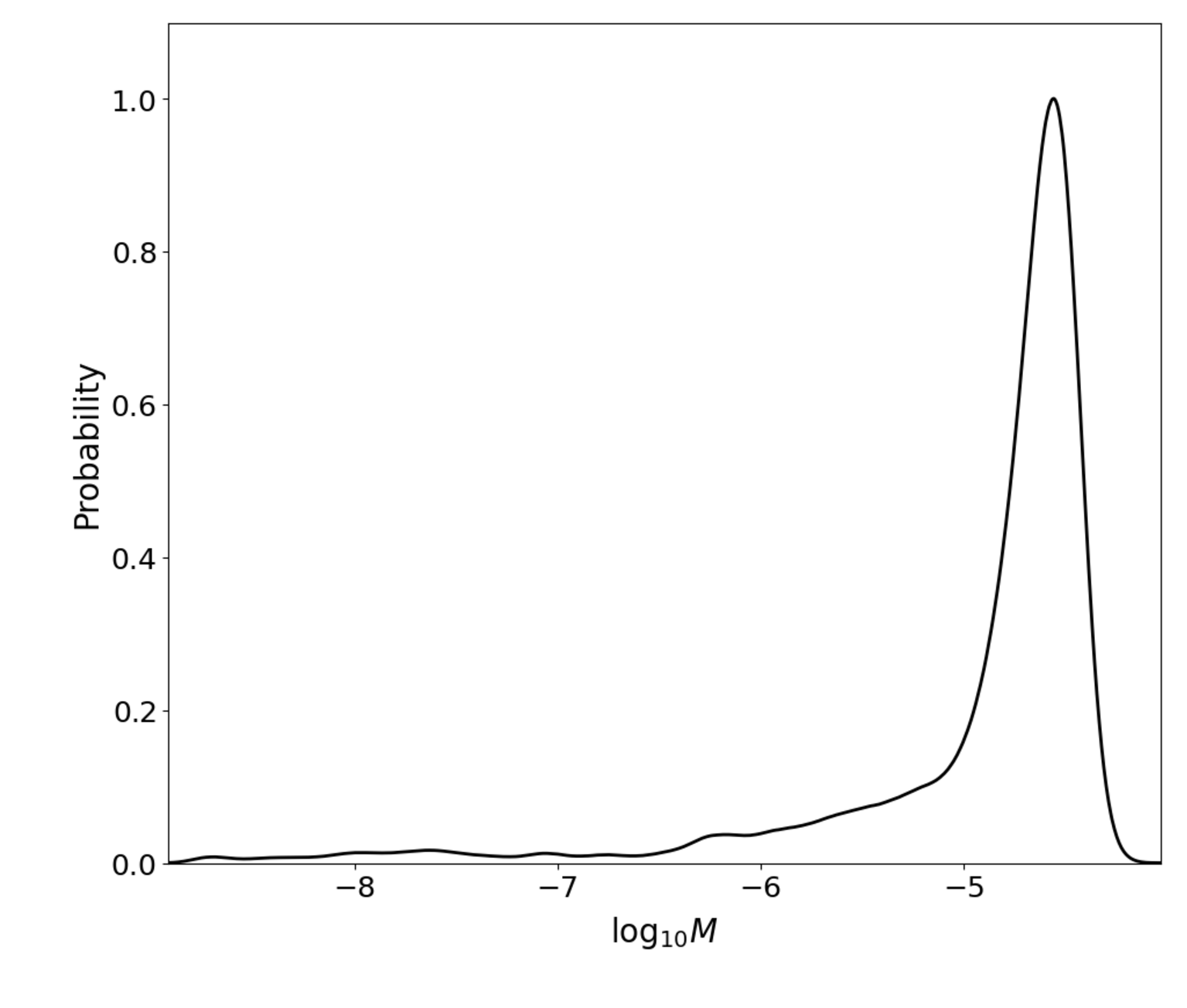}
}
\subfigure[]{
 \includegraphics[width=8cm, height = 5cm]{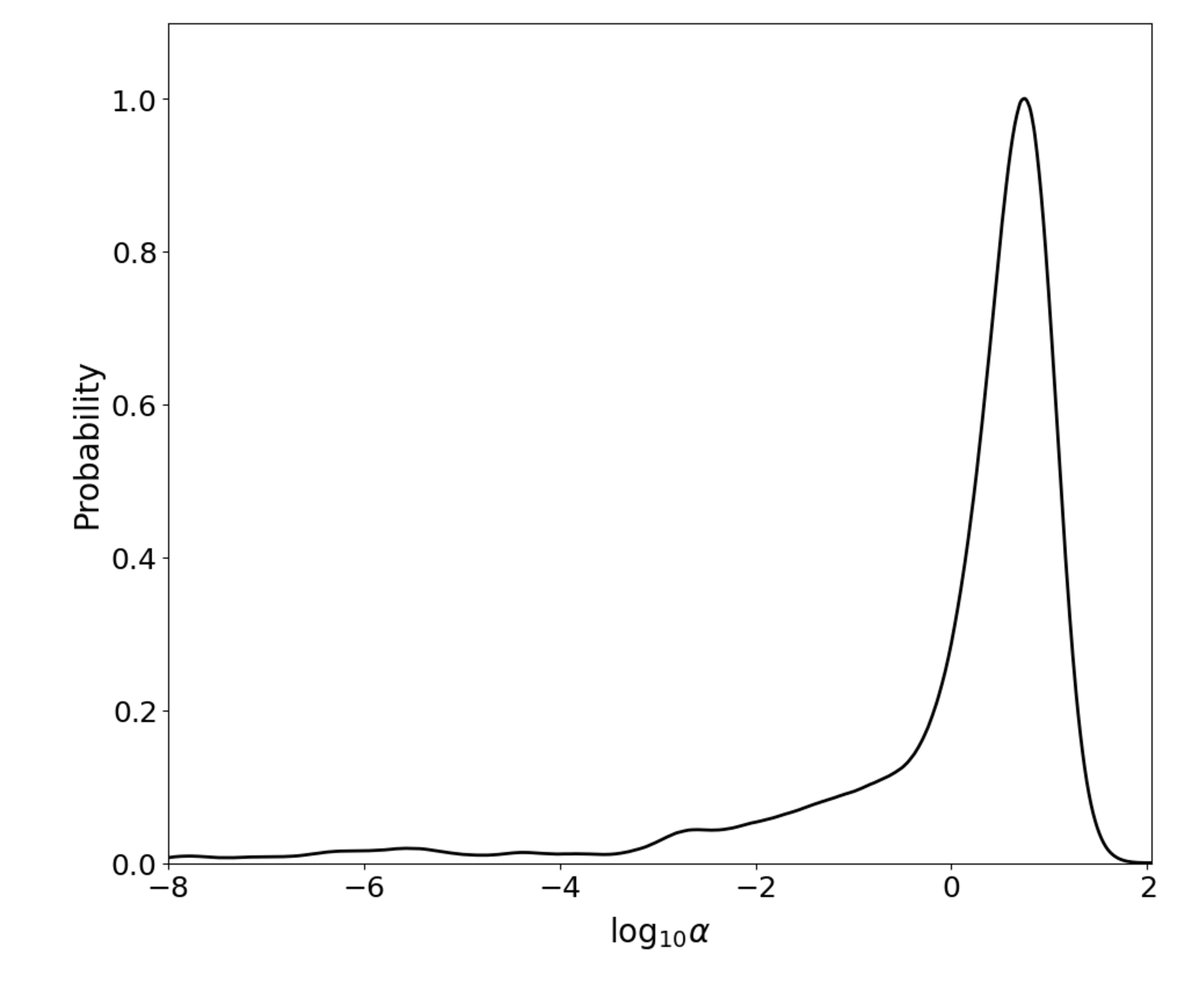}
\label{fig:alphabk}
}
\subfigure[]{
 \includegraphics[width=8cm, height = 5cm]{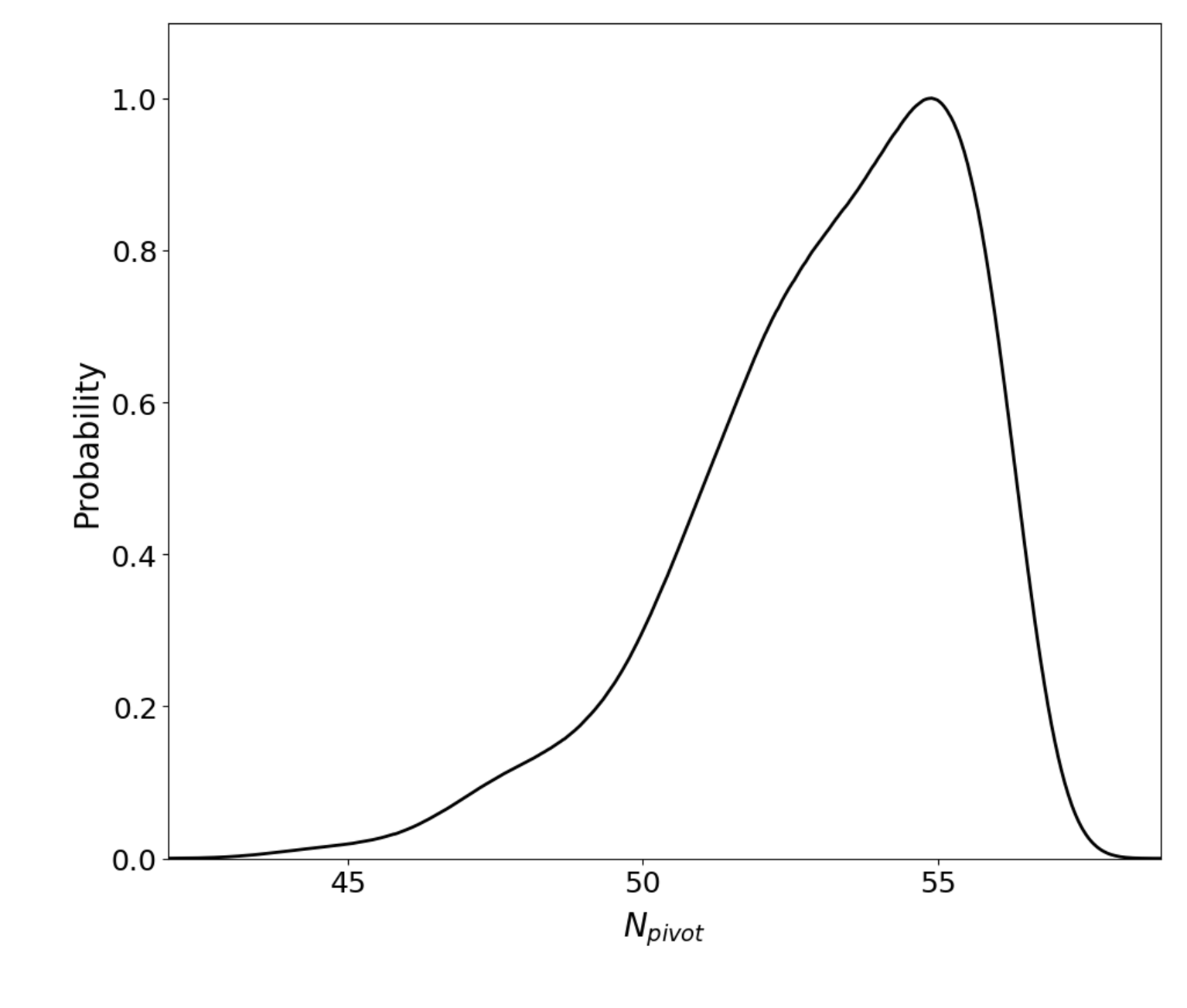}
}
\caption{Marginalized probability constraints on the potential parameters and $N_{pivot}$ from Planck-2018, BICEP/Keck (BK18) and BAO data}
 \label{fig:margconstr}
\end{center}
\end{figure}

\begin{figure}[h!]
\begin{center}
\subfigure[]{
 \includegraphics[width=8cm, height = 7cm]{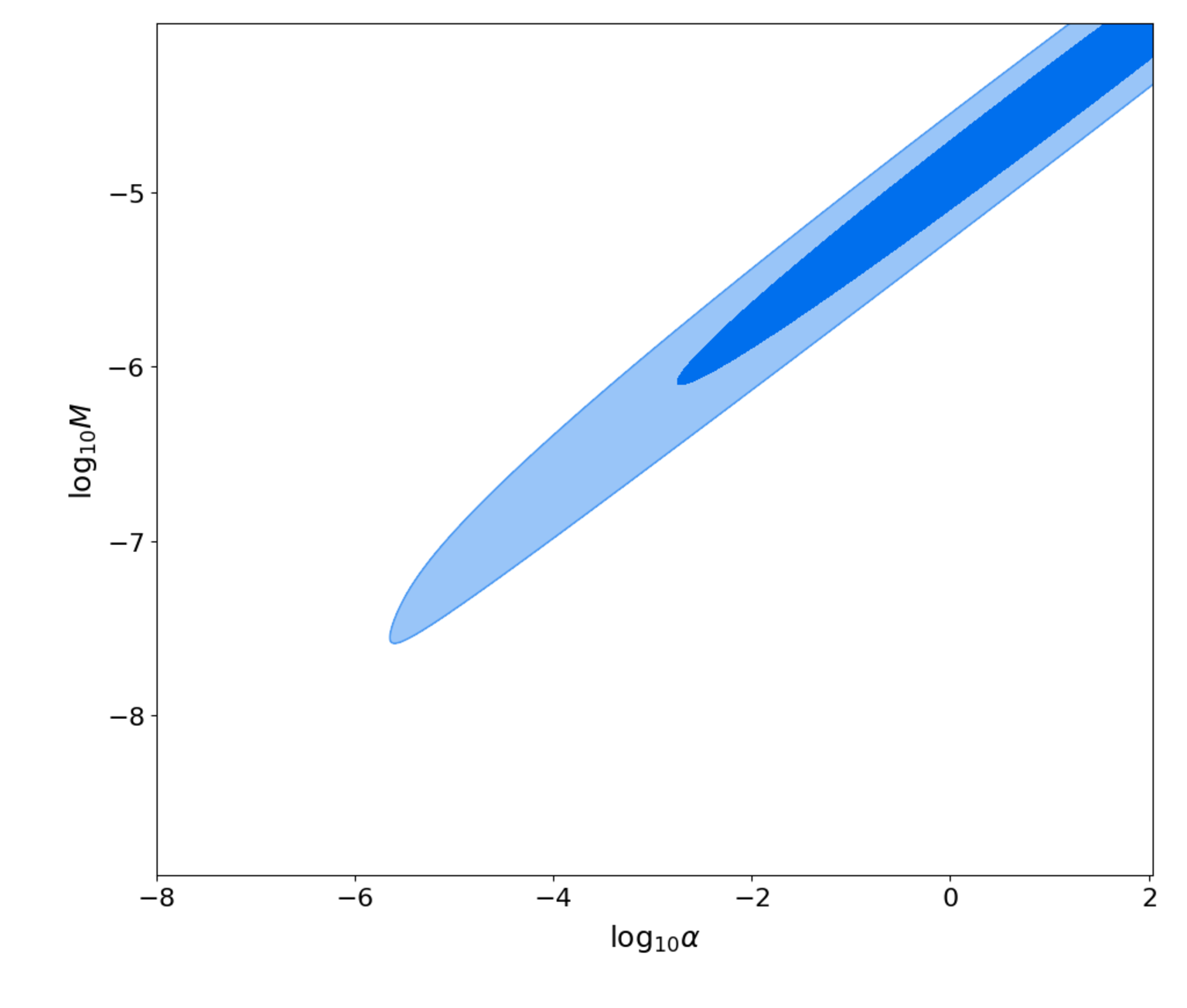}
 \label{fig:alpham}
}
\subfigure[ ]{
\includegraphics[width=7cm, height = 7cm]{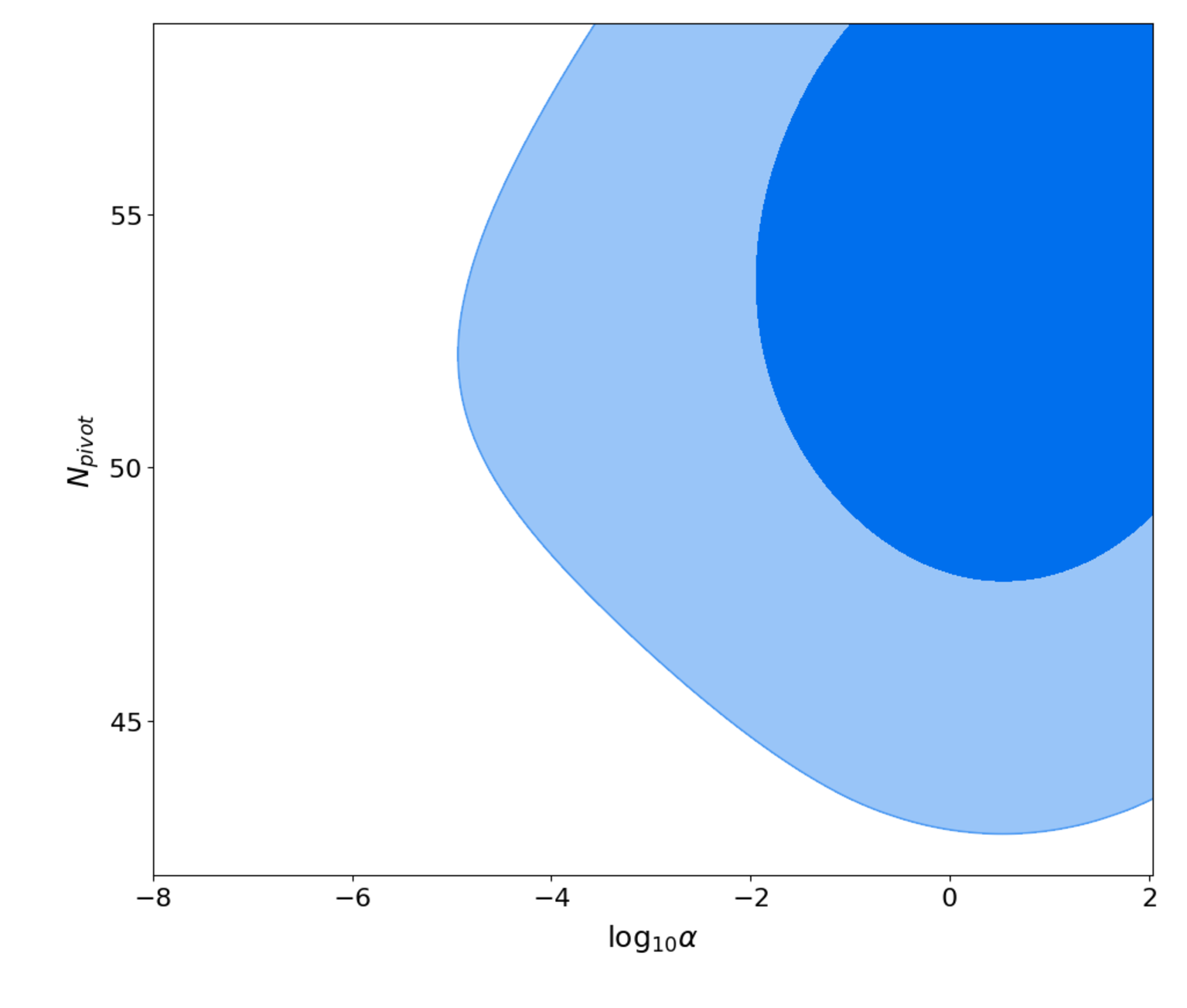}

 \label{fig:alphan}
}
\caption{Marginalized joint two-dimensional $68\%$ C.L. and $95\%$ C.L.
constraints on parameters of potential (a) and $N_{pivot}$ (b) using Planck-2018,
BICEP/Keck (BK18) and BAO data}
\label{fig:alphamandn}
\end{center}
\end{figure}

\begin{figure}[h!]
\begin{minipage}{0.6\textwidth}
\begin{center}
 \includegraphics[width=8cm, height = 7cm]{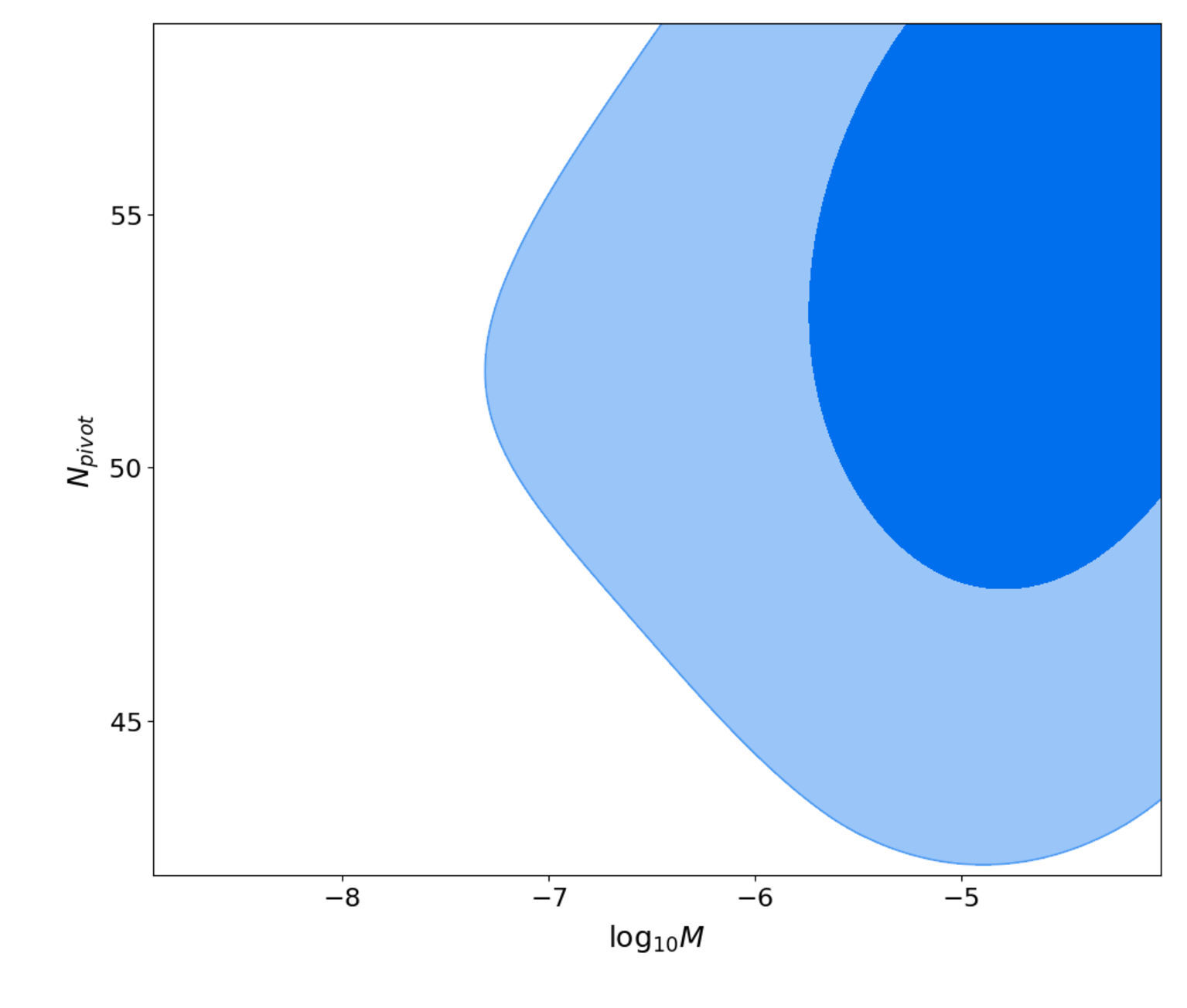}
\caption{$1\sigma$ and $2\sigma$ confidence contours in the  \\
potential parameter $\log_{10}M$ and $N_{pivot}$ parameter space }
\label{fig:mandn}
\end{center}
\end{minipage}%
\begin{minipage}{.5\textwidth}
\begin{center}
 \includegraphics[width=8.5cm, height = 7cm]{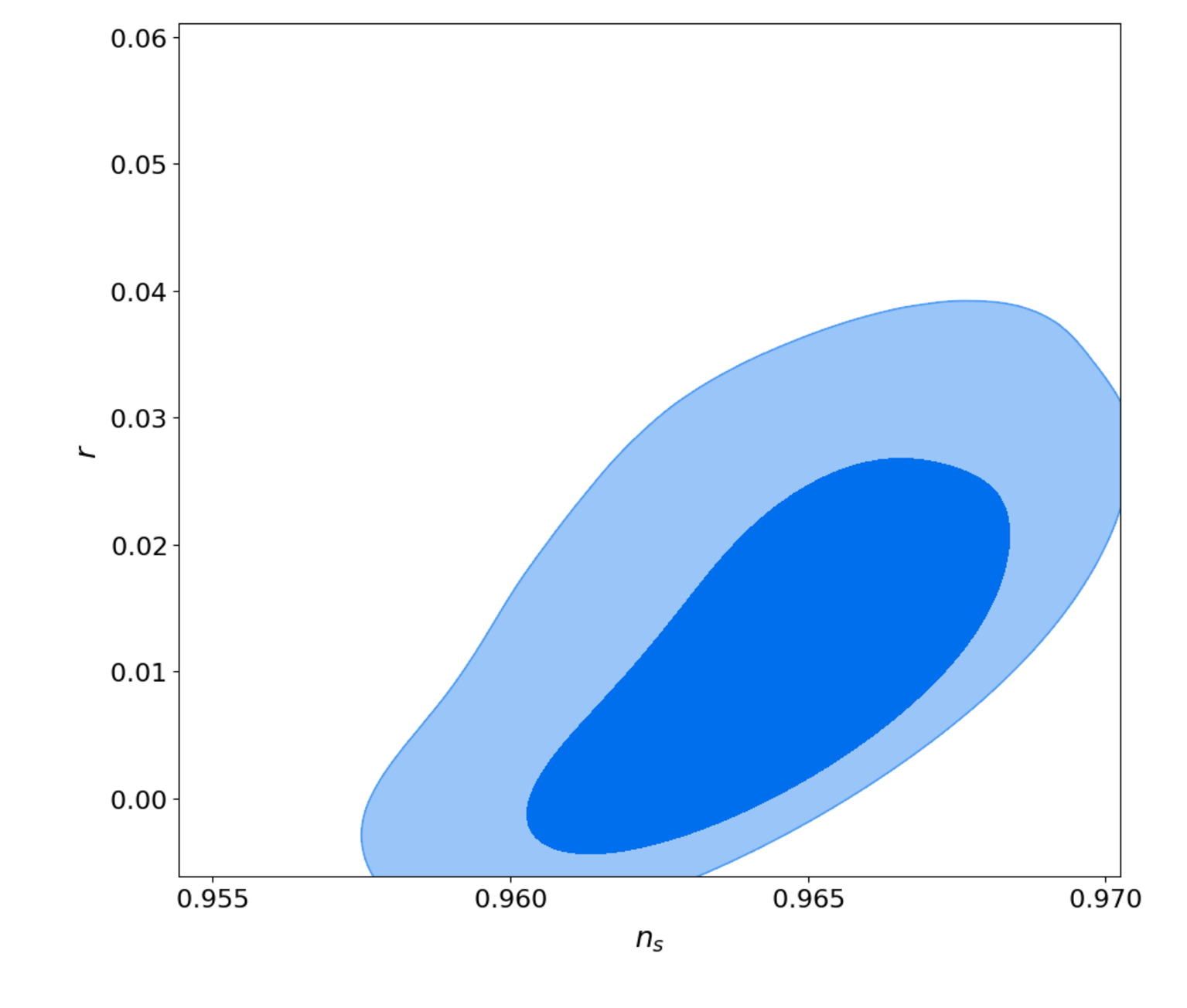}
\caption{1$\sigma$ and 2$\sigma$ confidence contours in the $n_s$-$r$  parameter  space}
\label{fig:nsr}
\end{center}
\end{minipage}
\end{figure}

The best fit values of scalar spectral index $n_s$ and tensor-to-scalar ratio $r$ for $\alpha$-Starobinsky  model, 
derived from the  potential parameters, lie within the  limits obtained by 
Planck-2018 and BICEP/Keck (BK18) observations as 
shown in Table \ref{Tab:constraint}. 
Fig.\ref{fig:margconstr} shows the marginalized probability distributions  for various inflationary parameters, while 
Figs. \ref{fig:alphamandn} and \ref{fig:mandn} illustrate the marginalized joint $68\%$ and $95\%$ constraints on the 
potential parameters $\alpha$ and $M$, as well as the e-folds $N_{pivot}$. It can be 
seen from Fig. \ref{fig:alphabk} that 
the most probable value of $\log_{10}\alpha$ is around $0.81$, however, it does not have any statistical significance as the 
limits are quite large compared to mean value. 
As depicted in Fig. \ref{fig:alpham} the potential parameters $\alpha$ and $M$ are strongly   correlated. 
However, Fig.\ref{fig:alphan} and Fig.\ref{fig:mandn} show that there is no correlation between the e-folds $N_{pivot}$ and 
potential parameters $\alpha$ and $M$ respectively. The joint $68\%$ and $95\%$ C.L. constraints on scalar spectral index $n_s$
and tensor-to-scalar ratio $r$ are shown in  Fig. \ref{fig:nsr}. As these two parameters are derived parameters, the constraints
on them are obtained from the constraints on potential parameters and $N_{pivot}$. 

\section{Conclusions} \label{conclusions}
 Starobinsky inflation \cite{Starobinsky:1980te,Starobinsky:1983zz} is one of the 
best suited model of inflation from the Planck-2018 \cite{Planck:2018jri} and 
BICEP/Keck (BK18) \cite{BICEP:2021xfz} observations  as it 
predicts smaller value for tensor-to-scalar ratio $r$. The inflaton potential 
of this model in the Einstein frame can be derived from no-scale supergravity 
with an underlying noncompact $SU(2,1)/SU(2)\times U(1)$ symmetry 
\cite{Ellis:2013xoa}, where the 
modulus field $T$ of the K\"ahler potential is fixed by other dynamics and 
inflaton field $\phi$ is a part of the chiral superfield with a minimum Wess-Zumino 
superpotential. By generalizing the coefficient of the 
logarithm of the volume
modulus field in the K\"ahler potential that parameterizes 
$SU(2,1)/SU(2)\times U(1)$ coset K\"ahler manifold, 
and considering a superpotential having both the volume modulus field and
inflaton field $\phi$, one can obtain a potential for inflaton that is similar to 
Starobinsky model with  a parameter $\alpha$ in the 
exponential \cite{Ellis:2013nxa}, known as $\alpha$-Starobinsky inflation. 
One can also obtain similar potential in
supergravity models with inflaton as a part of vector multiplet rather than chiral
multiplet \cite{Ferrara:2013rsa,Kallosh:2013yoa}, and the $\alpha$-Starobinsky model 
belongs to a class of superconformal inflationary models known as $E$-models 
of $\alpha$-attractors. 
 
 In this work we have used the inflation potential (\ref{alphapot}) to 
analyze $\alpha-$ attractor $E$-model, in the light of Planck-2018 and 
BICEP/Keck (BK18) \cite{BICEP:2021xfz} CMB observations and other LSS observations. 
We have used ModeChord, an updated version of ModeCode \cite{Mortonson:2010er}  to numerically 
compute the power spectra of both scalar and tensor perturbations without slow-roll 
approximation. 
This helps us to constrain the number of e-foldings $N_{pivot}$ and the model 
parameter $\alpha$ and $M$ using CMB and LSS observations by MCMC analysis using 
CosmoMC \cite{Lewis:2002ah}. To explore the parameter range we vary 
$log_{10}\alpha$ between $-8$ to $4$. In our analysis  we find that  
$log_{10}\alpha = 0.0^{+1.6}_{-5.6},\, \, \, 95\%$ C.L. from Planck-2018,
BICEP/Keck (BK18) and BAO data. The larger limits than the mean value on $\alpha$ indicates 
that the current observations are consistent with $\alpha=1$, which highlights the 
challenge in precisely determining $\alpha$ from current observational data, 
The $95\%$ upper limit on $\alpha$, $\log_{10}\alpha \le 1.6$ or $\alpha \le 39.81$, 
obtained in our analysis is slightly larger than the 
one obtained by Planck-2018 \cite{Planck:2018jri} i.e. $\log_{10}\alpha \le 1.3$, 
but,  smaller than the one obtained by \cite{Ellis:2021kad}, which is  
$\alpha\le 46$.   
By considering general reheating scenario we find that the number of e-foldings  
from the end of inflation to the time
when the length scales corresponding to $k_{pivot}$ leave the Hubble radius during
inflation $N_{pivot} = 53.2^{+3.9}_{-5.0}$ with $95\%$ C. L., which is sufficient
to solve horizon problem.
We also find that the potential parameters $M$ and $\alpha$ are strongly 
correlated, Fig.~\ref{fig:alpham};
and there is no correlation between these two parameters and $N_{pivot}$
 as shown in Fig.~\ref{fig:alphan} and Fig.~\ref{fig:mandn}.  

 The parameter $\alpha$ of the $\alpha$-Starobinsky model is inversely related 
to the curvature of the K\"ahler manifold. 
As the parameter $\alpha$ affects the amplitude of tensor perturbations, it will 
be possible with the future observations like CMB-S4 \cite{Abazajian:2019eic} and 
LiteBIRD \cite{Hazumi:2019lys} to obtain its 
precise value. The $\alpha$-Starobinsky inflation can 
be incorporated in a flipped $SU(5)\times U(1)$ GUT 
\cite{Ellis:2019jha,Ellis:2019opr}, where inflaton decay during
reheating can produce cold dark matter and also can have implications for neutrinos
masses and leptogenesis. The precise determination of the parameter $\alpha$ 
will help us in connecting the models of particle physics phenomenology with inflation
in the framework of supergravity and string theory.

\section{ACKNOWLEDGEMENTS}
The authors would like to thank ISRO Department of Space Govt. of India to provide financial
support via RESPOND programme Grant No. DS\_2B-13012(2)/47/2018-Sec.II.

\end{document}